\documentstyle[twocolumn,seceq]{jpsj}

\title{Resonant X-Ray Scattering from CeB$_6$}

\def\runtitle{Resonant X-Ray Scattering from CeB$_6$}
\def\runauthor{Jun-ichi {\sc Igarashi} and Tatsuya {\sc Nagao}}

\author{Jun-ichi {\sc Igarashi}\footnote{E-mail: jigarash@spring8.or.jp}
    and Tatsuya {\sc Nagao}$^{1}$}

\inst{Synchrotron Radiation Research Center, Japan Atomic Energy Research Institute, 
Mikazuki, Sayo, Hyogo  679-5148 \\
$^{1}$Faculty of Engineering, Gunma University, Kiryu, Gunma 376-8515}

\recdate{ February 25, 2002 }

\kword{resonant X-ray scattering, CeB$_{6}$, 
       orbital ordering, Ce $L_{{\rm III}}$ absorption edge, 
       magnetic form factor}

\abst{
We calculate the resonant x-ray  scattering (RXS) spectra
near the Ce $L_{\rm III}$ absorption edge in CeB$_6$,
on the basis of a microscopic model that the $4f$ states of Ce are
atomic while the $5d$ states form an energy band with a reasonable
density of states. 
In the initial state, we employ an effective Hamiltonian 
of Shiina {\it et al}. in the antiferro-quadrupole
(AFQ) ordering phase, while we construct the wave function consistent with 
the neutron scattering experiment in the magnetic ground state.
In the intermediate state,
we take full account of the intra-atomic Coulomb interaction.
Without assuming any lattice distortion, we obtain sufficient RXS
intensities on the AFQ superlattice spot. 
We obtain the spectral shape, the temperature and magnetic field 
dependences in good agreement with the experiment,
thus demonstrating the mechanism that the intensity 
is brought about by the modulation of $5d$ states through the anisotropic 
term of the $5d$-$4f$ Coulomb interaction. 
In the magnetic ground state, a small pre-edge peak is found 
by the $E_2$ process. On the magnetic superlattice spot,
we get a finite but considerably small intensity.
The magnetic form factor is briefly discussed.
}

\begin{document}
\sloppy
\maketitle
\input epsf.sty

\def\runtitle{Resonant X-Ray Scattering from CeB$_{6}$}
\def\runauthor{Jun-ichi {\sc Igarashi} and Tatsuya {\sc Nagao}}
\def\s{\sigma}
\def\l{\ell}
\def\ncr{\nonumber\\}


\section{Introduction}
Resonant x-ray scattering (RXS) has recently attracted much interest
as a useful tool to investigate the orbital order, which neutron scattering
experiments are usually difficult to probe.
The resonant enhancement for the prohibited Bragg reflection
corresponding to the orbital order has been observed in several
transition-metal compounds by using synchrotron radiation with photon energy 
around the $K$ absorption edge.
\cite{Murakami98a,Murakami98b,Murakami99b,Murakami99c,Murakami00b}

For such $K$-edge resonances, $4p$ states of transition metals are involved 
in the intermediate state in the electric dipolar ($E_1$) process, 
and they have to be modulated in accordance with the orbital order 
for observing signals.
At the early statge, for LaMnO$_3$, such a modulation was considered 
to come from the anisotropic term of the $4p$-$3d$ intra-atomic Coulomb 
interaction,\cite{Ishihara1,Ishihara2} 
but subsequent studies based on the band structure calculation
\cite{Elfimov,Benfatto,Takahashi1,Takahashi2} have revealed that
the modulation comes mainly from the lattice distortion via 
the oxygen potential on the neighboring sites.
Similar conclusions have been obtained for $t_{2g}$-electron systems,
such as YTiO$_3$\cite{Takahashi3} and YVO$_3$.\cite{Takahashi4}
This is because $4p$ states are so extending in space 
that they are very sensitive to the electronic structure at neighboring sites. 

Not only transition-metal compounds but also rare-earth-metal compounds
show the orbital order (usually an ordering of quadrupole moments). 
Recently, RXS experiments were carried out around the Ce $L_{\rm III}$ 
absorption edge in CeB$_6$, in which resonant enhancements
have been found  on the quadrupolar ordering superlattice spots.
\cite{Nakao1,Yakhou} In particular, Nakao {\em et al.}\cite{Nakao1}
have found a one-peak structure as a function of photon energy
on the spot ${\bf G}=(\frac{1}{2}\frac{1}{2}\frac{1}{2})$, 
which was assigned to the $E_1$ process.
Here ${\bf G}$ is the scattering vector in units of $2\pi/a$ with $a$
being the lattice constant.
They have also measured the temperature and the magnetic field dependences
of the intensities.
The purpose of this paper is to analyze their experimental result 
on the basis of a microscopic model and thereby to elucidate the mechanism
for RXS in CeB$_6$. Some of the results reported here were briefly
presented in a recent letter.\cite{Nagao}
In this paper, making a slight revision on the $5d$ density of states,
we describe explicitly the model as well as the calculational procedure.
We also add the calculation of the RXS spectra on the magnetic ground state.

\begin{figure}[t]
\centerline{\epsfxsize=7.50cm\epsfbox{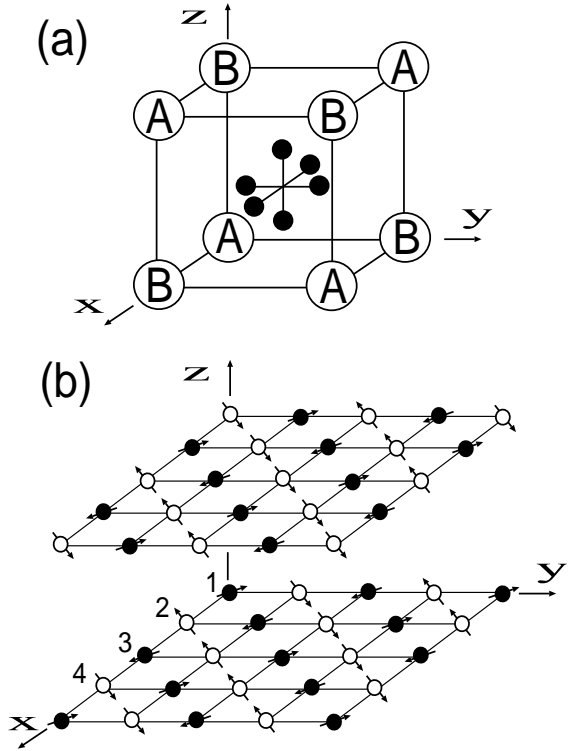}} 
\caption{
(a) Antiferro-quadrupolar structure of phase II.
Open circles are cerium atoms with A and B representing sublattices.
Solid small circles are boron atoms.
(b) Magnetic structure of phase III predicted by the neutron scattering
experiment. Only Ce atoms are shown.
The arrows indicate the direction of the magnetic moment.
The attached numbers represent sites of typical directions.
\label{fig.cryst}}
\end{figure}

Each Ce atom is considered to be in the $f^1$-configuration, $^2F_{5/2}$.
The $\Gamma_8$ quartet states have a lower energy than the $\Gamma_7$ doublet 
under the cubic crystal field. The $4f$ states are assumed to be atomic 
as a first approximation.
With decreasing temperatures the antiferro-quadrupole (AFQ) order appears
at $T_Q= 3.2$ K with an ordering wave vector 
${\rm Q}=(\frac{1}{2}\frac{1}{2}\frac{1}{2})$, as shown in 
Fig.~\ref{fig.cryst}(a). 
This phase transition originates from the intersite interaction
between the atomic $\Gamma_8$ states by lifting the degeneracy of 
the $\Gamma_8$ states. Ohkawa derived
an effective intersite interaction on the basis of a RKKY interaction.
\cite{Ohkawa1,Ohkawa2}.
Recently, Shiina {\it et} {\it al}.\cite{Shiina,Sakai,Thalmeier,Shiba} 
extended his model by taking full account of the symmetry of the interaction
as well as the order parameters. Thereby, they solved a longstanding
controversy between the neutron diffraction\cite{Effantin} and
NMR\cite{Takigawa} in the context of the induced order parameters
under the external magnetic field. We use the model Hamiltonian of
Shiina {\it et} {\it al}. within the mean field approximation (MFA)
for describing the initial state of the RXS process in the AFQ phase.

For $T<T_{\rm N}$ ($=2.4$ K), there appears a magnetic long-range order.
According to the neutron scattering experiment,\cite{Effantin} 
the ordering pattern of the magnetic moment is as shown 
in Fig.~\ref{fig.cryst}(b).
We construct the wave function in the magnetic ground state to be 
consistent with the ordering pattern, and use it as the initial state
of the RXS process.
Our calculation is limited to $T=0$, since an intersite interaction correctly 
reproducing the magnetic phase has not been derived yet.

In the intermediate state of the $E_1$ process, the $5d$ states of Ce 
are involved so that they have to be modulated in accordance with 
the superlattice spots.
Since the $4f$ states are so localized in space that their coupling to lattice
is very small. Actually the lattice distortion associated with the AFQ order
has not been observed. Therefore, it is highly possible that the modulation 
is brought about by the Coulomb interaction between the $5d$ states 
and the orbital ordering $4f$ states.
Introducing a reasonable density of states (DOS) for the $5d$ states,
we solve a scattering problem of the photo-excited $5d$ electron;
the scatterer is a complex of a $2p$ hole and a $4f$ electron. 

Using the solution in the intermediate state, and combining it to the result
in the initial state, we calculate the RXS spectra on an AFQ superlattice spot
${\bf G}=(\frac{1}{2}\frac{1}{2}\frac{1}{2})$
and on a magnetic superlattice spot 
${\bf G}=(\frac{1}{4}\frac{1}{4}\frac{1}{2})$ 
in the magnetic ground state.
In the AFQ phase, we obtain sufficient intensities on the AFQ spot
without assuming any lattice distortions, thereby demonstrating the mechanism
that the RXS intensity is brought about by the modulation of the $5d$ states 
in accordance with the AFQ order 
through the intra-atomic Coulomb interaction.
This situation is different from that of transition-metal compounds,
where there exists a sizable lattice distortion which is the primary
origin of the RXS spectra.
The calculated temperature and magnetic field dependences reproduce well 
the experiment.\cite{Nakao1}
We find that the azimuthal-angle dependence is closely related to
the symmetry of the AFQ order, independent of the details of model.
In the magnetic ground state, we find 
the RXS spectra on the AFQ superlattice spot 
as a smooth extension from the AFQ phase,
suggesting that the magnetic order has little influence on this spot.
We find a small pre-edge peak in the electric quadrupolar ($E_2$) process
there.\cite{Com1} On the magnetic superlattice spot,
we obtain a finite intensity, but it is about two order of magnitude 
smaller than that on the AFQ spot.

This paper is organized as follows.
In Sec.~II, we formulate the RXS spectra as a second order optical process.
In Sec.~III and IV, we describe the initial and the intemediate states
of RXS.
In Sec.~V, we discuss the calculated RXS spectra in comparison with 
experiments. Section VI is devoted to concluding remarks. 

\section{Formulation for Resonant X-Ray Scattering}

\begin{figure}[b]
\centerline{\epsfxsize=7.50cm\epsfbox{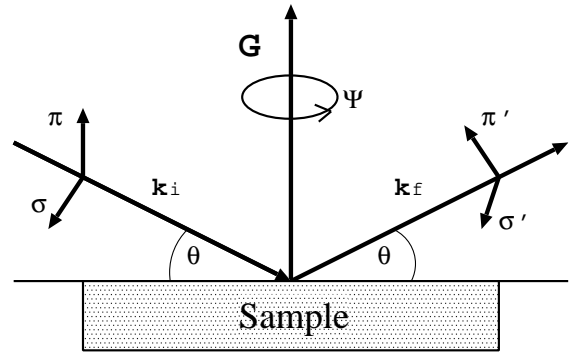}} 
\caption{
Scattering geometry of x-ray scattering. Incident photon with
wave vector ${\bf k}_i$ and polarization $\sigma$ or $\pi$ 
is scattered into the state with wave vector ${\bf k}_f$ and
polarization $\sigma'$ or $\pi'$ at Bragg angle $\theta$.
The sample crystal is rotated by azimuthal angle $\psi$
around the scattering vector ${\bf G}={\bf k}_f-{\bf k}_i$.
\label{fig.geom}}
\end{figure}
The conventional RXS geometry is shown in Fig.~\ref{fig.geom};
photon with frequency $\omega$, momentum ${\bf k}_i$ and polarization $\mu$ 
($=\sigma$ or $\pi$) is scattered into the state
with momentum ${\bf k}_f$ and polarization $\mu'$ ($=\sigma'$ or $\pi'$). 
The scattering vector is defined by ${\bf G}={\bf k}_f-{\bf k}_i$.
The cross section for the scattering vector ${\bf G}$ consists of three
terms:\cite{deBergevin,Blume1,Blume2}
\begin{eqnarray}
& &  \left. \frac{d\sigma}{d\Omega}\right|_{\mu\to\mu'} \propto 
 | J_{\mu\to\mu'}({\bf G},\omega)
 +\sum_{\alpha\alpha'}P'^{\mu'}_{\alpha}
    M_{\alpha\alpha'}({\bf G},\omega)P^{\mu}_{\alpha'}
\nonumber \\
& & + \sum_{\gamma\gamma'}Q'^{\mu'}_{\gamma}
         N_{\gamma\gamma'}({\bf G},\omega)Q^{\mu}_{\gamma'}
 |^2, 
\label{eq.cross}
\end{eqnarray}
where 
\begin{eqnarray}
 J_{\mu\to\mu'}({\bf G},\omega) &=&
 -\frac{{\rm i}\hbar\omega}{mc^2}\frac{1}{\sqrt{N}}\sum_j
  \exp(-{\rm i}{\bf G}\cdot{\bf r}_j) \nonumber \\
& \times &
\left(\frac{1}{2}{\bf L}({\bf G},j)\cdot {\bf A}''
  +{\bf S}({\bf G},j)\cdot{\bf B}\right),
\label{eq.nonresonance}
\\
 M_{\alpha\alpha'}({\bf G},\omega) &=& \frac{1}{\sqrt{N}}
  \sum_j\sum_{n,\Lambda}p_n(j)
       \exp(-{\rm i}{\bf G}\cdot{\bf r}_j) m \omega_{\Lambda n}^2 \nonumber \\
& \times &
  \frac{\langle\psi_n(j)|x_\alpha(j)|\Lambda\rangle
  \langle \Lambda|x_{\alpha'}(j)|\psi_n(j)\rangle}
       {\hbar\omega-(E_{\Lambda}-E_n(j))+i\Gamma},
\label{eq.dipole}
\\
 N_{\gamma\gamma'}({\bf G},\omega) &=& \frac{1}{\sqrt{N}}
  \sum_j \frac{k^2}{12}\sum_{n\Lambda'}p_n(j)
   \exp(-{\rm i}{\bf G}\cdot{\bf r}_j) \nonumber \\
& \times &
  \frac{m\omega_{\Lambda'n}^2 \langle \psi_n(j)|z_\gamma(j)|\Lambda'\rangle
  \langle\Lambda' |z_{\gamma'}(j)|\psi_n(j)\rangle}
   {\hbar\omega-(E_{\Lambda'}-E_n(j))+i\Gamma}.
\label{eq.quadrupole}
\end{eqnarray}
Here $c$ is the velocity of photon and $m$ is the electron mass.
Note that the cross section is an order of the number of Ce sites.

The first term (eq.~(\ref{eq.nonresonance})) represents a {\em non-resonant} 
term for the magnetic superlattice spots.
Since the magnetic moment comes mainly from $4f$ states which are well
localized, it may be a good approximation to assign the moment to each site.
The ${\bf L}({\bf G},j)$ and ${\bf S}({\bf G},j)$
are the form factors of the orbital and spin angular momenta at site $j$,
which are given by\cite{Blume2,Trammell}
\begin{eqnarray}
 {\bf L}({\bf G},j) &=& \frac{1}{2}\sum_n p_n(j)
\nonumber \\
\hspace*{-0.3cm}
& \times &
       \langle\psi_n(j)|f(-{\bf G}\cdot{\bf r})\mbox{\boldmath$\ell$}
       +\mbox{\boldmath$\ell$}f(-{\bf G}\cdot{\bf r})|\psi_n(j)\rangle ,
\label{eq.orbital}\\
 {\bf S}({\bf G},j) &=& \sum_n p_n(j)
  \langle\psi_n(j)|{\rm e}^{-{\rm i}{\bf G}\cdot{\bf r}}{\bf s}
  |\psi_n(j)\rangle,
\label{eq.spin}
\end{eqnarray}
with
\begin{equation}
 f(x) = 2\sum_{m=0}^{\infty} \frac{({\rm i}x)^m}{(m+2)m!}.
\end{equation}
Here $|\psi_n(j)\rangle$ represents the initial state at site $j$,
which will be evaluated within the MFA discussed in the next section. 
We take the thermal average with probability $p_n(j)$.
Operators $\mbox{\boldmath$\ell$}$ and ${\bf s}$ represent 
the orbital and spin angular 
momenta with the center of site $j$.
The ${\bf L}({\bf G},j)$ and ${\bf S}({\bf G},j)$ converge to the local orbital 
momentum and the spin angular momentum with ${\bf G}\to 0$.
The scattering geometry is contained in quantities ${\bf A}''$ and ${\bf B}$,
which are defined by
\begin{eqnarray}
 {\bf A}''&=& {\bf A}'-({\bf A}'\cdot{\bf\hat G}){\bf\hat G},
  \quad {\bf A}'=-4\sin^2\theta(\hat{\mbox{\boldmath $\epsilon$}}'\times
  \hat{\mbox{\boldmath $\epsilon$}}),\\
 {\bf B} &=& \hat{\mbox{\boldmath $\epsilon$}}'\times
 \hat{\mbox{\boldmath $\epsilon$}}
             + ({\bf\hat k}_f\times \hat{\mbox{\boldmath $\epsilon$}}')
               ({\bf\hat k}_f\cdot \hat{\mbox{\boldmath $\epsilon$}})
             - ({\bf\hat k}_i\times \hat{\mbox{\boldmath $\epsilon$}})
               ({\bf\hat k}_i\cdot \hat{\mbox{\boldmath $\epsilon$}}')
\nonumber \\
& - & ({\bf\hat k}_f\times \hat{\mbox{\boldmath $\epsilon$}}')
        \times ({\bf\hat k}_i\times \hat{\mbox{\boldmath $\epsilon$}}),
\end{eqnarray}
where $\hat{\mbox{\boldmath $\epsilon$}}$ and
$\hat{\mbox{\boldmath $\epsilon$}}'$ are the initial
and scattered polarizations, and
${\bf\hat k}_i$, ${\bf\hat k}_f$, and ${\bf\hat G}$ are normalized
vectors of ${\bf k}_i$, ${\bf k}_f$, and ${\bf G}$.

The second term in eq.~(\ref{eq.cross}) describes a {\em resonant} term
by the $E_1$ process,
where an electron in $2p$ states is virtually excited to $5d$ states and
subsequently is recombined with the core hole.
Since the $2p$ states are well localized around Ce sites,
it is a good approximation to describe the scattering tensor
as a sum of the contribution from each site of the core hole.
The initial state $|\psi_n(j)\rangle$ at site j has an energy $E_n(j)$.
The intermediate state $|\Lambda\rangle$ consists of an excited electron
on $5d$ states and a hole on $2p$ states with energy $E_{\Lambda}$.
$\omega_{\Lambda n}=(E_{\Lambda} - E_{n})/\hbar$.
The life-time broadening width $\Gamma$ of the core hole is assumed 
to be $2$ eV.
The dipole operators $x_\alpha(j)$'s are defined as
$x_1(j)=x$, $x_2(j)=y$, and $x_3(j)=z$ in the coordinate frame fixed 
to the crystal axes with the origin located at the center of site $j$.
The dipole matrix element 
$A_{dp}=\langle 5d|r|2p\rangle =
   \int_0^{\infty} R_{5d}(r)rR_{2p}(r)r^2{\rm d}r $
is implicitly included as a square in the expression
($R_{5d}(r)$ and $R_{2p}(r)$ are the radial wave functions 
for the $5d$ and $2p$ states, respectively).
It is estimated as $A_{dp}=3.67\times 10^{-11}$ cm for a Ce$^{3+}$ atom
within the Hartree-Fork (HF) approximation.\cite{Cowan}
The $P^\mu$ and $P'^{\mu}$ are geometrical factors for the incident 
and scattered photons, respectively, which are explicitly written
in the Appendix of ref.~31.

The third term in eq.~(\ref{eq.cross}) describes a {\em resonant} term
by the $E_2$ process, where an electron in $2p$ states is virtually 
excited to $4f$ states and subsequently is recombined with the core hole.
In eq.~(\ref{eq.quadrupole}), $k$ is the wavenumber of the incident 
(and scattered) photon, which is $\sim 2.91\times 10^8$ cm$^{-1}$
around the $L_{\rm III}$ edge.
The intermediate states $|\Lambda'\rangle$ consist of an excited electron 
on $4f$ states and a hole in $2p$ states with
energy $E_{\Lambda'}$. Quadrupole operators are defined as
$z_1\equiv(\sqrt{3}/2)(x^2-y^2)$, $z_2\equiv(1/2)(3z^2-r^2)$, 
$z_3\equiv\sqrt{3}yz$, $z_4\equiv\sqrt{3}zx$, and $z_5\equiv\sqrt{3}xy$ 
in the coordinate frame fixed to the crystal axes.
The quadrupole matrix element 
$A_{fp} = \langle 4f|r^2|2p\rangle =
   \int_0^{\infty} R_{4f}(r)r^2R_{2p}(r)r^2{\rm d}r$ is included as a square
in the expression ($R_{4f}(r)$ is the $4f$ radial wave function).
It is estimated as $A_{fp}=5.69\times 10^{-20}$ cm$^2$ 
for a Ce$^{3+}$ atom within the HF approximation.\cite{Cowan}
The $Q^\mu$ and $Q'^{\mu'}$ are geometrical factors for the incident and
scattered photons, respectively, which are explicitly written in the Appendix
of ref.~31.

\section{Initial State} 

Each Ce atom is approximately in the $4f^1$-configuration, 
$^2 F_{5/2}$, in CeB$_6$.
The $4f$ states are so localized that their wave functions are well described
to be atomic in the HF approximation. The cubic crystal field lifts
the degeneracy; quadruply degenerate $\Gamma_8$ states have a lower energy than
the doubly degenerate $\Gamma_7$ states.
Since the $\Gamma_8$-$\Gamma_7$ separation energy is estimated as large as
$\sim 540$ K, it is sufficient to consider only the $\Gamma_8$ states 
in the description of the initial state, which are explicitly written as
\begin{eqnarray}
   |+\uparrow\rangle
 &=& \sqrt{\frac{5}{6}}\left|+\frac{5}{2} \right\rangle
    +\sqrt{\frac{1}{6}}\left|-\frac{3}{2} \right\rangle,\nonumber\\
   |+\downarrow\rangle 
 &=& \sqrt{\frac{5}{6}}\left|-\frac{5}{2} \right\rangle
    +\sqrt{\frac{1}{6}}\left|+\frac{3}{2} \right\rangle,\nonumber\\
   |-\uparrow   \rangle &=& \left|+\frac{1}{2} \right\rangle,\nonumber\\
   |-\downarrow \rangle &=& \left|-\frac{1}{2} \right\rangle.
\end{eqnarray}
The state $|M\rangle$ has the $z$-component $M$ of the total angular momentum, 
where the $z$ axis is taken along the $[0,0,1]$ direction.  
Symbols $\tau$($=\pm$) and $\sigma$($=\uparrow,\downarrow$)
in $|\tau,\sigma\rangle$ represent non-Kramers' and Kramers' pairs, 
respectively.

\subsection{Quadrupolar Ordering Phase ($T_{\rm N} < T < T_{\rm Q}$)}

Some quadrupolar ordering is established (Phase II)
below $T_{\rm Q}=3.2$ K. An effective Hamiltonian, which is derived on 
the basis of the RKKY interaction, is known to work well:\cite{Shiba}
\begin{eqnarray}
\hat H & = & D \sum_{\langle i,j \rangle}
\left[ (1+\delta)
  \mbox{\boldmath$\mu$}_{i} \cdot \mbox{\boldmath$\mu$}_{j}
  + \tau_{i}^{y} \tau_{j}^{y}
  + \epsilon \mbox{\boldmath$\sigma$}_{i} \cdot \mbox{\boldmath$\sigma$}_{j}
\right.  \nonumber \\
& & \hspace*{0.80cm} \left.
  + \frac{1+\epsilon}{2} 
   (  \mbox{\boldmath$\tau$}_{i} ' \cdot \mbox{\boldmath$\tau$}_{j} '
    + \mbox{\boldmath$\eta$}_{i} \cdot \mbox{\boldmath$\eta$}_{j}
    + \mbox{\boldmath$\zeta$}_{i} \cdot \mbox{\boldmath$\zeta$}_{j}
    ) \right] \nonumber \\
    & & \hspace*{0.80cm} + g \mu_{\rm B} 
   \sum_{i} \mbox{\boldmath$J$}_{i}\cdot \mbox{\boldmath$H$}, 
\label{eq.hamilton}
\end{eqnarray}
with
\begin{eqnarray}
   \mbox{\boldmath$J$} &=& \frac{7}{3}(\mbox{\boldmath$\sigma$}
                          +\frac{4}{7}\mbox{\boldmath$\eta$}), \\
 \mbox{\boldmath$\mu$} 
     &=& (2\tau^y\sigma^x,2\tau^y\sigma^y,2\tau^y\sigma^z), \\
 \mbox{\boldmath$\tau$}' &=& (\tau^z,\tau^x),\\
 \mbox{\boldmath$\eta$} &=& ((\sqrt{3}\tau^x-\tau^z)\sigma^x,
                (-\sqrt{3}\tau^x-\tau^z)\sigma^y,2\tau^z\sigma^z),\\
 \mbox{\boldmath$\zeta$} &=& ((-\sqrt{3}\tau^z-\tau^x)\sigma^x,
                ( \sqrt{3}\tau^z-\tau^x)\sigma^y,2\tau^x\sigma^z),
\end{eqnarray}
where $\langle i,j\rangle$ represents the sum over nearest neighboring 
Ce pairs. Operators ${\bf\tau}$ and ${\bf\sigma}$ represent the spin matrix 
acting on the variables $\tau$ and $\sigma$ of the state
$|\tau\sigma\rangle$, respectively.
This system has been extensively studied within the
MFA by Shiina et al;\cite{Shiina,Sakai,Thalmeier,Shiba} 
$\delta\sim 0.2$ and $\epsilon\sim 1$ are known to be suitable for CeB$_6$.
In the following, we simply summarize the result of the MFA 
in connection to the RXS spectra.

In the absence of the magnetic field, an AFQ order is set in,
as shown in Fig.~\ref{fig.cryst}(a). 
We have three types of possible ordered phase,
in which one of the staggered quadrupole moments, $\langle\tilde O_{xy}\rangle$
($\equiv 4\langle\tau_y\sigma_z\rangle$),
$\langle\tilde O_{yz}\rangle$
($\equiv 4\langle\tau_y\sigma_x\rangle$),
and $\langle\tilde O_{zx}\rangle$
($\equiv 4\langle\tau_y\sigma_y\rangle$), is finite.
Here $\langle X\rangle$ indicates the thermal average of operator $X$.
We simply call them as the $O_{xy}$, $O_{yz}$, and $O_{zx}$ phases.
For example, in the $O_{xy}$ phase, applying the MFA to 
eq.~(\ref{eq.hamilton}), we obtain the eigenstates of the Hamiltonian 
at site $j$:
\begin{eqnarray}
 |\psi_1(j)\rangle &=& \frac{1}{\sqrt{2}}\left\{ 
                 |+\uparrow\rangle + {\rm i}|-\uparrow\rangle \right\},\\
 |\psi_2(j)\rangle &=& \frac{1}{\sqrt{2}}\left\{ 
                 |+\downarrow\rangle - {\rm i}|-\downarrow\rangle \right\},\\
 |\psi_3(j)\rangle &=& \frac{1}{\sqrt{2}}\left\{ 
                 |+\uparrow\rangle - {\rm i}|-\uparrow\rangle \right\},\\
 |\psi_4(j)\rangle &=& \frac{1}{\sqrt{2}}\left\{ 
                 |+\downarrow\rangle + {\rm i}|-\downarrow\rangle \right\},
\end{eqnarray}
with the eigenvalues,
\begin{eqnarray}
 E_1(j)=E_2(j) &=& \mp zD(1+\delta)|4\langle\tau_y\sigma_z\rangle|, 
\label{eq.eneA}\\
 E_3(j)=E_4(j) &=& \pm zD(1+\delta)|4\langle\tau_y\sigma_z\rangle|, 
\label{eq.eneB}
\end{eqnarray}
where the upper(lower) sign in eqs.~(\ref{eq.eneA}) and (\ref{eq.eneB})
is for A(B) sublattice. The $4\langle\tau_y\sigma_z\rangle$ is 
the staggered order parameter self-consistently determined, 
and $z$ ($=6$) is the number of nearest neighboring pairs. 
Note that Kramers' pairs, $|\psi_1\rangle$ and $|\psi_2\rangle$, as well as
$|\psi_3\rangle$ and $|\psi_4\rangle$, are still degenerate in Phase II.
The probability $p_n(j)$ appeared in the preceding section is given by 
$\propto \exp(-E_n(j)/T)$. Note that
$p_1(j)(=p_2(j))>p_3(j)(=p_4(j))$ at A sublattice, and vice versa 
at B sublattice. 

\begin{figure}[t]
\centerline{\epsfxsize=7.50cm\epsfbox{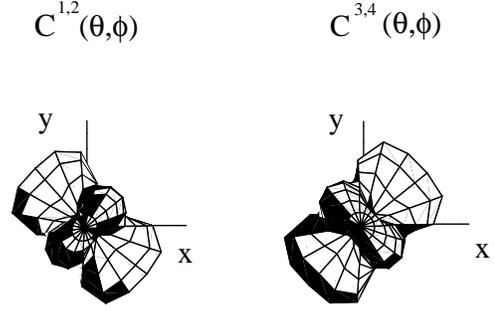}} 
\caption{
Charge distributions $C^{1,2}(\theta,\phi)$ and $C^{3,4}(\theta,\phi)$.
\label{fig.charge}}
\end{figure}

The angular dependences of the charge density for those states
are defined by
\begin{eqnarray}
 C^{1,2}(\theta,\phi) &=& \frac{1}{2 |R_{4f}(r)|^2} \sum_{n=1,2}
 |\langle r,\theta,\phi|\psi_n\rangle|^2, \\
 C^{3,4}(\theta,\phi) &=& \frac{1}{2 |R_{4f}(r)|^2} \sum_{n=3,4}
 |\langle r,\theta,\phi|\psi_n\rangle|^2. 
\end{eqnarray}
We calculate these quantities using the atomic function in the HF 
approximation.\cite{Cowan} As shown in Fig.~\ref{fig.charge},
$C^{1,2}(\theta,\phi)$ is along the $[1,-1,0]$ direction,
while $C^{3,4}(\theta,\phi)$ is along the $[1,1,0]$ direction.
The charge distribution after thermal average is along the $[1,-1,0]$ 
direction at A sublattice, since $p_1(j)(=p_2(j))>p_3(j)(=p_4(j))$ there.
Such anisotropy leads to a modulation in the $5d$ states
through the Coulomb interaction in the RXS process.

\begin{figure}[t]
\centerline{\epsfxsize=7.50cm\epsfbox{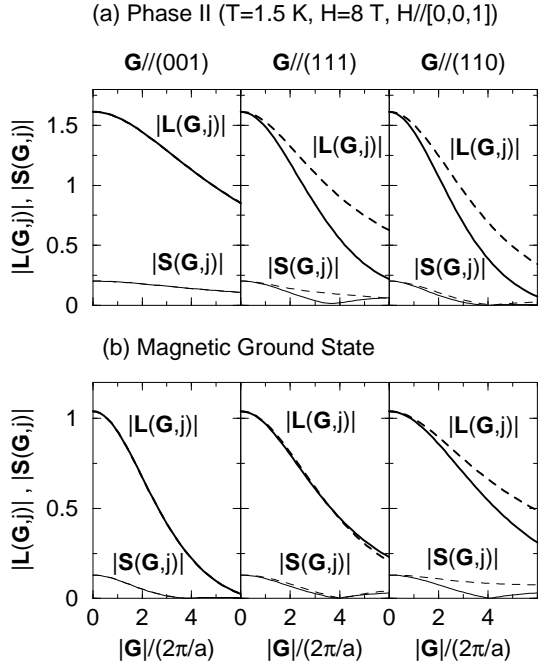}} 
\caption{
Absolute values of the orbital moment form factor, $|L({\bf G},j)|$, 
and of the spin moment form factor, $|S({\bf G},j)|$, 
as a function of $|{\bf G}|$($\equiv 4\pi\sin\theta/\lambda$).
(a) Phase II ($T=1.5$ K, $H=8$ T, ${\bf H}\parallel [0,0,1]$.
Thick solid and thick broken lines represent $|L({\bf G},j)|$
for site $j$ belonging to A and B sublattices, respectively.
Thin solid and thin broken lines are for $|S({\bf G},j)|$.
(b) Magnetic ground state. Thick solid and thick broken lines 
represent $|L({\bf G},j)|$ for site $j$ belonging to site 1 and 3,
and belonging to site 2 and 4, respectively.
Thin solid and thin broken lines are for $|S({\bf G},j)|$.
\label{fig.moment} }
\end{figure}

The magnetic field induces the staggered octupole moment
in addition to the staggered quadrupole moments,
as shown in the upper panel of Fig.~\ref{fig.spext}.
As shown in the same figure, the octupole moment has little influence
on the RXS spectra.
The magnetic field also induces the uniform dipole moment.
Recently Saitoh {\em et al.} have carried out a neutron diffraction 
experiment under magnetic field, and have reported the magnetic form factors.
\cite{Saitoh} 
They have argued that a considerable amount of the magnetic moment is
distributed around B atoms.
In this context, it may be instructive to evaluate the magnetic form factors
on Ce atoms, although they are not directly related to the present RXS study.
We use the atomic wave function within the HF approximation 
in  eqs.~(\ref{eq.orbital}) and (\ref{eq.spin}).\cite{Cowan}
Figure \ref{fig.moment}(a) shows the form factors on Ce atoms
for various values of ${\bf G}$.
$H=8$ T (${\bf H}\parallel [0,0,1]$), $T=1.5$ K.
The orbital moment is much larger than the spin moment,
and the form factors decrease monotonically with increasing values
of $|{\bf G}|$. These values do not fit the experimental data.
\cite{Saitoh,Hanzawa}

\subsection{Magnetic Phase ($T < T_{\rm N}$)}

With further decreasing temperature ($T<T_{\rm N}$), a magnetic long-range 
order appears (Phase III).  
The ordering pattern predicted by the neutron scattering experiment
is shown in Fig.~\ref{fig.cryst}(b). The magnetic moment ${\bf m}({\bf r}_j)$ 
is directing on the $ab$ plane; it is given by
\begin{eqnarray}
 {\bf m}({\bf r}_j) &\propto&
   \left\{\left[\cos\left({\bf k}_1\cdot{\bf r}_j+\frac{\pi}{4}\right)
     +\cos\left({\bf k'}_1\cdot{\bf r}_j-\frac{\pi}{4}\right)\right]
     {\bf u}_{{\bf k}_1} \right. \nonumber\\
    &+&
    \left. \left[\cos\left({\bf k}_2\cdot{\bf r}_j+3 \frac{\pi}{4}\right)
     +\cos\left({\bf k'}_2\cdot{\bf r}_j+\frac{\pi}{4}\right)\right]
      {\bf u}_{{\bf k}_2}\right\}, \nonumber \\
\label{eq.magdis}
\end{eqnarray}
where ${\bf u}_{{\bf k}_1}$ and ${\bf u}_{{\bf k}_2}$ are unit vectors
along the $[-1,1,0]$ direction and along the $[1,1,0]$ direction, respectively,
and ${\bf k}_1=(\frac{1}{4}\frac{1}{4}\frac{1}{2})$,
${\bf k}_2=(\frac{1}{4}\frac{\bar 1}{4}\frac{1}{2})$,
${\bf k'}_1=(\frac{1}{4}\frac{1}{4}0)$,
${\bf k'}_2=(\frac{1}{4}\frac{\bar 1}{4}0)$.
Of course, there must exist other magnetic domains in which the moments are
directing on the $bc$ and $ca$ planes.

The magnetic domain described by eq.~(\ref{eq.magdis}) is expected to come
from the $O_{xy}$ phase with splitting the degeneracy of Kramers' doublet.
We have no reliable effective inter-site interaction between Kramers' doublets;
the inter-site interaction given by eq.~(\ref{eq.hamilton})
cannot describe the magnetic state. Therefore we are satisfied to derive
the ground state wave function consistent with the distribution of 
the magnetic moment experimentally determined.
First we note that the angular momentum operator ${\bf J}$ along ${\bf n}$ 
($=(\cos\phi,\sin\phi,0)$)
is represented within the space of $|\psi_1\rangle$ and $|\psi_2\rangle$
as 
\begin{equation}
 {\bf n\cdot J}= A|\psi_1\rangle\langle\psi_2|
                  + A^*|\psi_2\rangle\langle\psi_1|,
\label{eq.magA}
\end{equation}
with $A=-(1/3)\exp(-{\rm i}\phi)-({\rm i}/\sqrt{3})\exp({\rm i}\phi)$,
while it is represented within the space of $|\psi_3\rangle$ 
and $|\psi_4\rangle$ as 
\begin{equation}
 {\bf n\cdot J}= B|\psi_3\rangle\langle\psi_4|
                  + B^*|\psi_4\rangle\langle\psi_3|,
\label{eq.magB}
\end{equation}
with $B=-(1/3)\exp(-{\rm i}\phi)+({\rm i}/\sqrt{3})\exp({\rm i}\phi)$.
Considering the magnetic ordering pattern in Fig.~\ref{fig.cryst}(b),
and noting that ${\bf J}$ is pointing to the direction opposite to the local
magnetic moment vector,
we seek the eigenstate with the negative eigenvalue for eq.~(\ref{eq.magA}) 
with $\phi=3\pi/4$ for site 1, eq.~(\ref{eq.magB}) with $\phi=-3\pi/4$ 
for site 2, eq.~(\ref{eq.magA}) with $\phi=-\pi/4$ for site 3, 
and eq.~(\ref{eq.magB}) with $\phi=\pi/4$ for site 4.
Thus we have the wave function in the ground state as
\begin{equation}
 |\psi_g\rangle = \left\{ 
\begin{array}{ll}
\frac{1}{\sqrt{2}}\left({\rm e}^{-{\rm i}3\pi/4}|\psi_1\rangle
                                              +|\psi_2\rangle\right),
& {\rm for\ site\ 1}, \\
 \frac{1}{\sqrt{2}}\left({\rm e}^{{\rm i}3\pi/4}|\psi_3\rangle
                                                +|\psi_4\rangle\right),
& {\rm for\ site\ 2}, \\
\frac{1}{\sqrt{2}}\left({\rm e}^{{\rm i}\pi/4}|\psi_1\rangle
                                               +|\psi_2\rangle\right),
& {\rm for\ site\ 3}, \\
\frac{1}{\sqrt{2}}\left({\rm e}^{-{\rm i}\pi/4}|\psi_3\rangle
                                               +|\psi_4\rangle\right),
& {\rm for\ site\ 4}, \\
\end{array}
\right.
\end{equation}
with the eigenvalue $-0.91$.
Considering the $g$-factor $6/7$, we have the local magnetic moment 
$0.78\mu_{\rm B}$, which is close to the value $0.66\mu_{\rm B}$
from the analysis of the $^{11}B$-NMR measurement.\cite{Tsuji}

The orbital and spin form factors are evaluated by using the above
wave functions.
Figure \ref{fig.moment}(b) shows the calculated result as a function of 
$|{\bf G}|$.
The form factors decreases monotonically with increasing values of $|{\bf G}|$,
which behavior is similar to that in Phase II under the magnetic field.
We need these quantities to evaluate the non-resonant term,
eq.~(\ref{eq.nonresonance}), in the RXS spectra.
However, as shown later, the non-resonant term gives
much smaller contribution than the resonant terms in the ground state.

\section{Intermediate State}

In the $E_1$ process, an electron is excited from $2p$ states 
to $5d$ states at a Ce site.
The $2p$-core hole states are split into 
the states with $j_p=3/2$ and $j_p=1/2$ ($j_p$ is the total angular momentum)
due to the strong spin-orbit interaction.
In the following, we consider only the $j_p=3/2$ states ($L_{\rm III}$ edge).

\begin{figure}[t]
\centerline{\epsfxsize=7.50cm\epsfbox{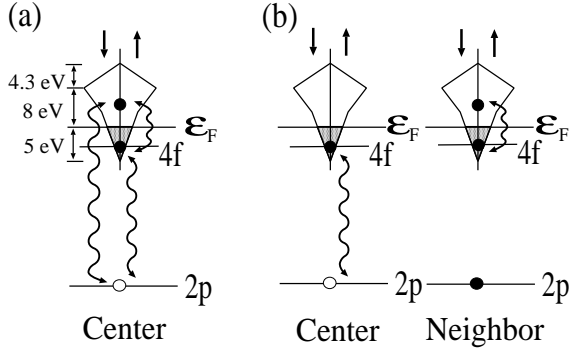}} 
\caption{
Schematic view of the configuration of the intermediate state of the $E_1$
process:
(a) the $5d$ excited electron is at the central site where the core hole 
is sitting;
(b) the $5d$ excited electron is at the neighboring sites.
The $5d$-DOS is schematically shown on the upper part.
Electrons are occupied below the Fermi level.
The wavy lines represent the Coulomb interaction taken into account.
Arrows indicate spins of $5d$ states.
\label{fig.level} }
\end{figure}

Different from the $4f$ states, the $5d$ states are rather extended in space,
so that they form an energy band with width $\sim 15$ eV
through the hybridization with boron $p$ states. 
We use a shape shown in Fig.~\ref{fig.level} as a model of
the $5d$ density of states (DOS).\cite{Com2} 
We disregard the dependence on the symmetries $xy$, $yz$, $zx$, $x^2-y^2$, 
and $3z^2-r^2$.
It is assumed to be occupied by one electron 
per Ce site in the initial state.
Such a rather arbitrary choice of the $5d$-DOS may be justified in
a semi-quantitative study,
since the RXS spectra is not sensitive to the shape and the filling of the
$5d$-DOS. The retarded Green's function for the excited $5d$ electron
is defined by
\begin{equation}
 G^{5d}(\hbar\omega)
   = \int_{\epsilon_F}^{\infty} 
   \frac{\rho^{5d}(\epsilon)}
        {\hbar\omega-\epsilon+{\rm i}\delta}{\rm d}{\epsilon},
\label{eq.5dgreen}
\end{equation}
where $\rho^{5d}(\epsilon)$ is the $5d$-DOS, and $\epsilon_F$
is the Fermi energy.\cite{Com2}
Note that the energy of $5d$ states included implicitly an average
interaction with electrons in $4f$ states.

Now we consider the resolvent $1/(\hbar\omega-H_{\rm int})$ with $H_{\rm int}$
being the Hamiltonian 
in the configuration that there are one excited electron in the $5d$ band, 
one electron per Ce site in the $4f$ states, and one hole in $2p$ states 
at the {\em central} site (Fig.~\ref{fig.level}(a)).
We neglect the screening effect by the already occupying electrons in the $5d$
band. When the excited ``$5d$ electron" is away from the central site, 
it can freely move; the Green's function, eq.~(\ref{eq.5dgreen}),
describes this motion.
At the same time, the ``$4f$ electron" is interacting with
the ``$2p$ hole" on the central site (Fig.~\ref{fig.level}(b)). 
First, considering the Coulomb interaction between the $4f$ electron 
and the $2p$ hole and also the spin-orbit interaction at the central site, 
we solve the eigenvalue problem for the complex of the $4f$ electron 
and the $2p$ hole. In this calculation, we use the Slater integrals 
evaluated by the HF approximation in a Ce$^{3+}$ 
atom, which are listed in Table \ref{table1}.\cite{Com3}
Let $|\lambda\rangle$ be the eigenstate with energy $E_{\lambda}$.
On the other hand, when the $5d$ electron comes onto the central site, 
it interacts with the $2p$ hole as well as the $4f$ electron.
We treat this system as a scattering problem of the $5d$ electron,
in which the scatterer has 56 ($4\times 14$) internal degrees of freedom 
specified by $\lambda$ on the central site. 
Taking account of the multiple scattering of the $5d$ electron 
by this scatterer, 
we obtain the expression of the resolvent at the central site as
\begin{eqnarray}
& & \left(\frac{1}{\hbar\omega - H_{\rm int} + 
{\rm i}\delta}\right)_{m^ds^d\lambda;m'^ds'^d\lambda'}
\nonumber \\
&  = &[G^{5d}(\hbar\omega+{\rm i}\Gamma-E_\lambda)^{-1}
     \delta_{\lambda\lambda'}\delta_{m^dm'^d}\delta_{s^ds'^d} 
  \nonumber\\ 
     &-& V_{m^ds^d\lambda;m'^ds'^d\lambda'}]^{-1},
\label{eq.matrix}
\end{eqnarray}
where $(m^ds^d)$ specifies a state of the $5d$ electron.
Matrix $V_{m^ds^d\lambda;m'^ds'^d\lambda'}$ includes the Coulomb
interaction between the $5d$ electron and the $2p$ hole and between the $5d$ 
electron and the $4f$ electron. We have to exclude the average interaction
energy between the $5d$ and $4f$ electrons from this matrix,
since it is already included into
 the energy of the $5d$ band. 
The right hand side of eq.~(\ref{eq.matrix}) is a matrix 
with dimensions $560\times 560$ ($560=10\times 56$), 
which we numerically invert.

\begin{table}[t]
\caption{Slater integrals and the spin-orbit interaction
for Ce$^{3+}$ atoms in 
the Hartree-Fock approximation (in units of eV).}
\begin{center}
\begin{tabular}{llll}
\hline
\hline
 $F^{k}(4f,4f)$ &  $F^{k}(2p,5d)$ & $F^{k}(2p,4f)$ & $F^{k}(4f,5d)$  \\
$F^0$ \hspace*{0.2cm}26.08 & 
$F^0$ \hspace*{0.2cm}15.58 &
$F^0$ \hspace*{0.2cm}37.68 & 
$F^0$ \hspace*{0.2cm}13.92 \\
$F^2$ \hspace*{0.2cm}12.43 & 
$F^2$ \hspace*{0.2cm}0.568 & 
$F^2$ \hspace*{0.2cm}1.540 &
$F^2$ \hspace*{0.2cm}3.810\\
$F^4$ \hspace*{0.2cm}7.807 &
          & 
          & 
$F^4$ \hspace*{0.2cm}1.894 \\
$F^6$ \hspace*{0.2cm}5.618 & 
          & 
          &
         \\
\hline
  & $G^{k}(2p,5d)$ & $G^{k}(2p,4f)$ & $G^{k}(4f,5d)$  \\
                            & 
$G^1$ \hspace*{0.2cm}0.485 & 
$G^2$ \hspace*{0.2cm}0.144 & 
$G^1$ \hspace*{0.2cm}1.633 \\
                            & 
$G^3$ \hspace*{0.2cm}0.286 & 
$G^4$ \hspace*{0.2cm}0.093 & 
$G^3$ \hspace*{0.2cm}1.397 \\
                            &
                            &  
                            &
$G^5$ \hspace*{0.2cm}1.086 \\
\hline
  $\zeta_{4f}=$ 0.132 & $\zeta_{5d}=$ 0.138 &   & \\
\hline
\hline
\end{tabular} 
\end{center}
$\ast$In the RXS calculation, the above values of the anisotropic terms 
are reduced by multiplying a factor 0.8, while the values for 
$F^{0}(nl,n'l')$ are replaced by much smaller values,
$F^{0}(4f,5d) = 3.0$, $F^{0}(4f,4f) = 7.0$,
$F^{0}(2p,5d) = 4.0$ and $F^{0}(2p,4f) = 12.0$. 
\label{table1}
\end{table}
Once we obtain the resolvent, we can calculate the scattering amplitude,
eq.~(\ref{eq.dipole}), by using the relation,
\begin{eqnarray}
& &  \sum_{\Lambda}\frac{|\Lambda\rangle\langle\Lambda|}
       {\hbar\omega-(E_{\Lambda}-E_n(j))+{\rm i}\Gamma} \nonumber \\
& =& \sum_{m^ds^d\lambda}\sum_{m'^ds'^d\lambda'} \nonumber \\
& &
|m^ds^d\lambda\rangle
 \left(\frac{1}{\hbar\omega - H_{\rm int} + {\rm i}\delta}\right)_
{m^ds^d\lambda;m'^ds'^d\lambda'}
  \langle m'^ds'^d\lambda'|, \nonumber \\
\end{eqnarray}
for $j$ at the central site.
It should be noted here that this resolvent is the same at all sites 
of core hole. The scattering amplitudes become different at different
sublattices after multiplying the matrix elements 
of the dipole operators between the initial and the intermediate states.
Using the resolvent, we can also calculate the absorption coefficient 
$A(\omega)$ in the $E_1$ process,
\begin{eqnarray}
 A(\omega) & \propto & \sum_j \sum_{n\alpha}
   p_n(j)\langle\psi_n(j)|x^\alpha|m^ds^d\lambda\rangle 
\nonumber \\
& \times &
   \left(-\frac{1}{\pi}\right)
 {\rm Im}\left(\frac{1}{\hbar\omega - H_{\rm int} + {\rm i}\delta}\right)_
    {m^ds^d\lambda;m'^ds'^d\lambda'} \nonumber\\
  &\times& \langle m'^ds'^d\lambda'|x^\alpha|\psi_n(j)\rangle ,
\label{eq.absorp}
\end{eqnarray}
where ${\rm Im}X$ indicates the imaginary part of the quantity $X$.

In the $E_2$ process, an electron is excited from $2p$ states to $4f$ 
states at Ce sites.
Two electrons occupy the $4f$ states on the Ce site.
Since the $4f$ states are well localized, it may be sufficient
to consider only the core-hole site for the intermediate state 
$|\Lambda'\rangle$ in eq.~(\ref{eq.quadrupole}).
We obtain $|\Lambda'\rangle$ by numerically diagonalizing 
the Hamiltonian matrix within the space of two $4f$ electrons and one $2p$ 
hole. We fully take account of the multiplets, where the necessary Slater 
integrals as well as the spin-orbit interaction parameter are listed in
Table \ref{table1}.\cite{Com3}

\section{RXS Spectra}

\subsection{Quadrupolar Ordering Phase ($T_{\rm N}<T<T_{\rm Q}$)}

\begin{figure}[t]
\centerline{\epsfxsize=7.50cm\epsfbox{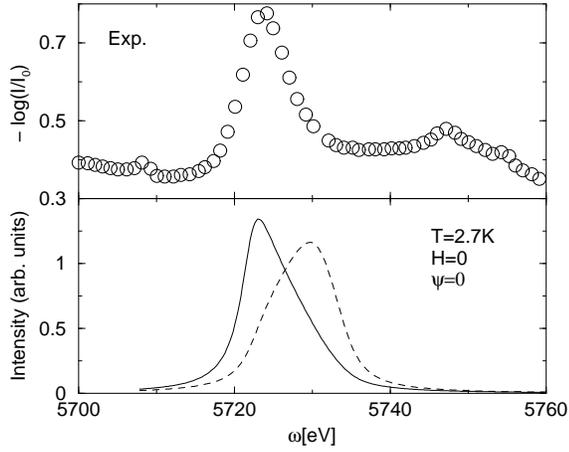}} 
\caption{
Absorption coefficient $A(\omega)$ of the Ce $L_{\rm III}$ edge
in comparison with the experimental spectra (ref.~37).
$T=2.7$ K and $H=0$.
Solid and broken lines represent the spectra calculated with 
$F^0(2p,5d)=3$ eV and 0, respectively.
\label{fig.absorp}}
\end{figure}

\begin{figure}[t]
\centerline{\epsfxsize=7.50cm\epsfbox{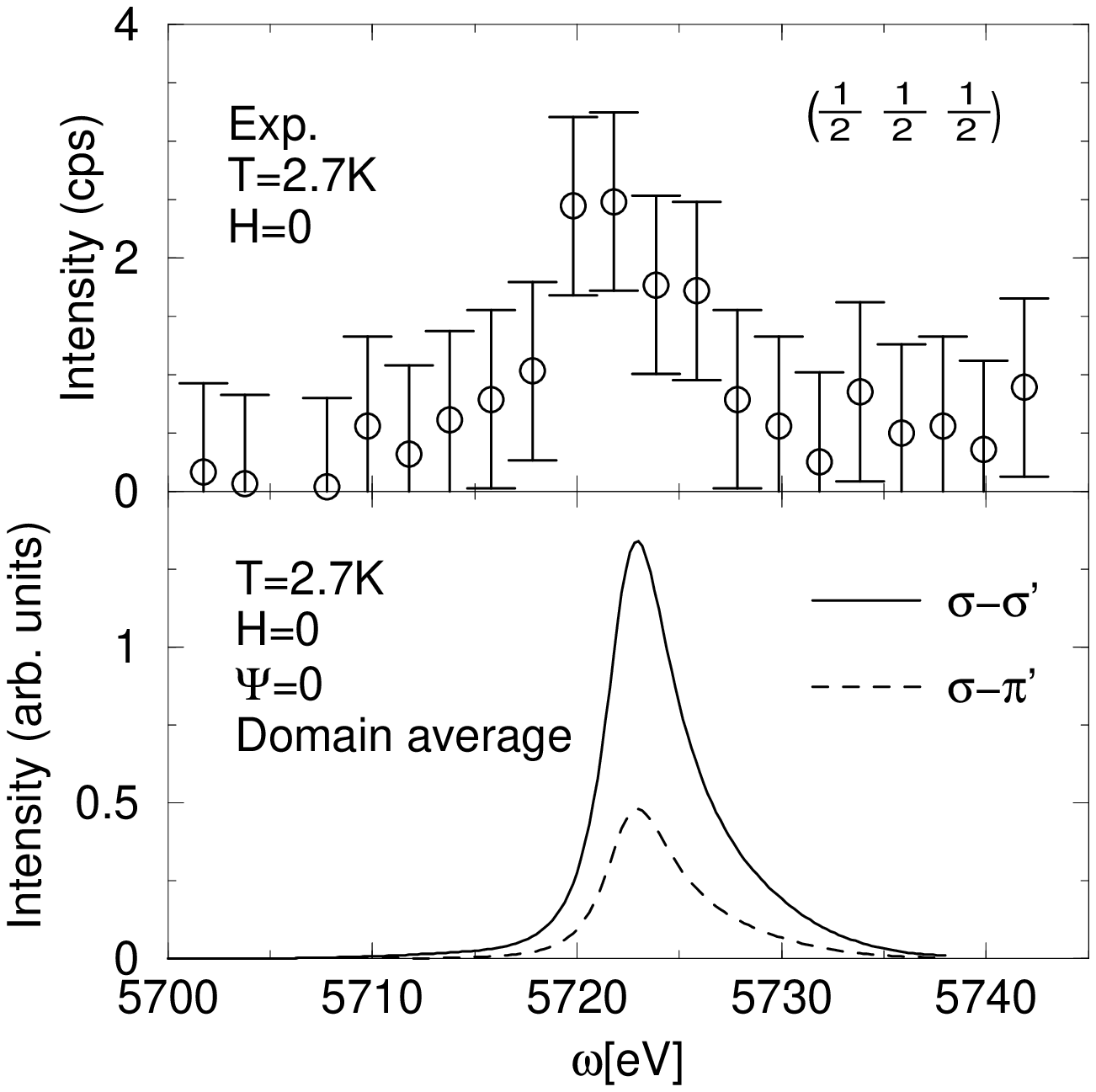}} 
\caption{
RXS spectra for ${\bf G}=(\frac{1}{2}\frac{1}{2}\frac{1}{2})$ 
in comparison with the experiment (ref.~14). 
$T=2.7$ K, $H=0$, and $\psi=0$. 
Intensities from three domains are averaged with equal weight.
The solid and broken lines represent the intensities
for the $\sigma\to\sigma'$ channel and
the $\sigma\to\pi'$ channel, respectively.
\label{fig.spene}}
\vskip 20pt
\centerline{\epsfxsize=7.50cm\epsfbox{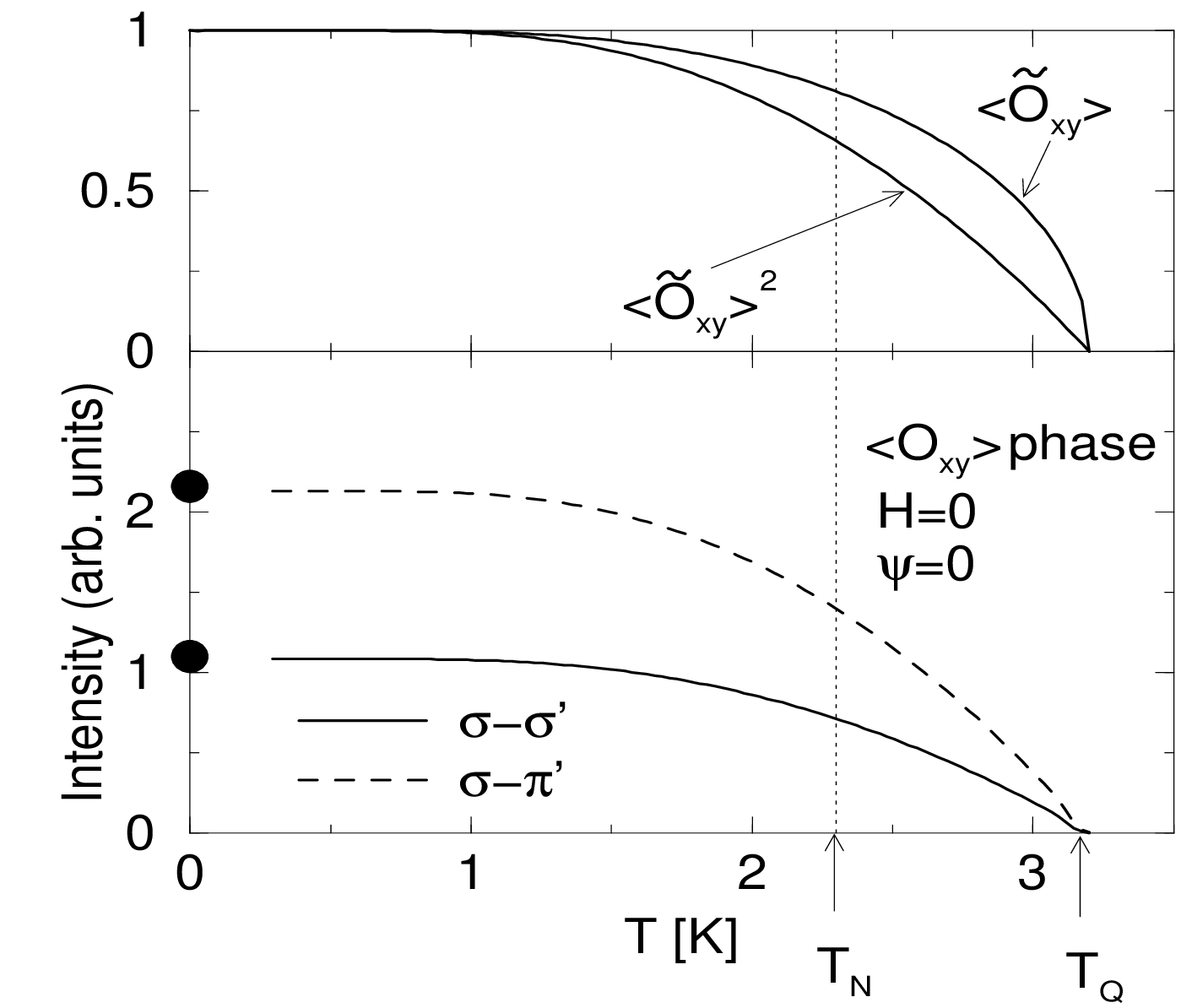}} 
\caption{
RXS intensity of the peak at $\hbar\omega=5722$ eV 
for ${\bf G}=(\frac{1}{2}\frac{1}{2}\frac{1}{2})$ in the $O_{xy}$ phase,
as a function of temperature.
$H=0$ and $\psi=0$.
The solid and broken lines represent the intensities for
the $\sigma\to\sigma'$ channel and the $\sigma\to\pi'$ channel, respectively.
Solid circles at $T=0$ is the peak intensity evaluated in the magnetic 
ground state.
The upper panel shows the staggered quadrupole moment
$\langle\tilde O _{xy}\rangle$ 
and its square $\langle\tilde O _{xy}\rangle^2$ in the $O_{xy}$ phase
within the MFA.
\label{fig.sptemp}}
\end{figure}

Before going to the discussion of the RXS spectra, 
we first calculate the absorption coefficient $A(\omega)$ 
with the help of eq.~(\ref{eq.absorp}). 
The contribution from the $E_2$ process is safely neglected, 
since it is too small by the estimate of the HF transition matrix elements.
Figure \ref{fig.absorp} shows the calculated result at $T=2.7$ K, $H=0$,
in comparison with the experiment.\cite{Nakao2}
The core-hole energy is adjusted such that the peak is located 
at $\hbar\omega=5722$ eV. Note that the temperature dependence is negligible.  
The spectrum is a reflection of the $5d$-DOS; 
the attractive interaction between the $5d$ electron and the core hole
makes the peak move to the low-energy region.
The calculated spectrum corresponds well to the experimental shape around
the $L_{\rm III}$ edge, indicating the appropriateness of the assumed $5d$ DOS.
Considerable intensities in the high energy region in the experimental spectra
may come from the mixing of $5d$ states to other states such as 
$6s$ states of Ce and $3s$ states of B.
Now we calculate the RXS spectra for 
${\bf G}=(\frac{1}{2}\frac{1}{2}\frac{1}{2})$,
following the procedure in the preceding section.
Figure \ref{fig.spene} shows the calculated spectra as a function 
of photon energy, in comparison with the experiment.\cite{Nakao1}
$T=2.7$ K and $H=0$.
The azimuthal angle $\psi$ is set to be zero
such that the scattering plane contains the $[1,-1,0]$ crystal axis.
The relative volumes of three domains, the $O_{xy}$, $O_{yz}$ and $O_{zx}$
phases, are not known in the experiment,
so that we have tentatively averaged the contributions from three phases 
with equal weight.
Since the dependence on the photon energy is the same in three domains,
the spectral shape is not influenced by the change of the relative volume
of domains. Only changeable are the relative intensities between 
the $\sigma\to\sigma'$ channel and the $\sigma\to\pi'$ channel;
in the $O_{xy}$ phase, the intensity of the $\sigma\to\pi'$ channel
is larger than that of the $\sigma\to\sigma'$ channel.
Such polarization analysis has not been done in the experiment.\cite{Nakao1}
We obtain an one-peak structure from the $E_1$ process, in good agreement
with the experiment.\cite{Nakao1} 
The contribution of the $E_2$ process is two order of magnitude less than
that of the $E_1$ process, so that we have no pre-edge peak visible.

\begin{figure}
\centerline{\epsfxsize=7.50cm\epsfbox{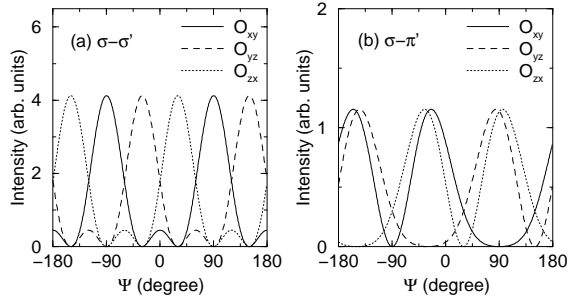}} 
\caption{
RXS intensity of the peak at $\hbar\omega=5722$ eV
for ${\bf G}=(\frac{1}{2}\frac{1}{2}\frac{1}{2})$, 
as a function of azimuthal angle $\psi$. $T=2.7$ K and $H=0$.
(a) the $\sigma\to\sigma'$ channel;
(b) the $\sigma\to\pi'$ channel.
The solid, broken, and dotted lines represent the contributions
from the $O_{xy}$, $O_{yz}$, and $O_{zx}$ phases, respectively.
\label{fig.spazim}}
\end{figure}
\noindent
As mentioned before,
the intensity arises from the $5d$ states modulated by the anisotropy
in the $4f$ charge distribution through the intra-atomic $5d$-$4f$ Coulomb 
interaction in the intermediate state. 
Figure \ref{fig.sptemp} shows the intensity of the peak at $\hbar\omega=5722$
eV in the $O_{xy}$ phase, as a function of temperature.
The curves are extended to $T < T_{\rm N}$ by assuming the $O_{xy}$ phase.
Its temperature dependence seems similar to that of 
$\langle\tilde O_{xy}\rangle^2$, indicating a direct reflection of
the AFQ order.

Another important quantity is the dependence on the azimuthal angle.
Figure \ref{fig.spazim} shows the RXS intensity of the peak 
at $\hbar\omega=5722$ eV.
The contributions from three domains are separately shown.
The dependence of the $\sigma\to\sigma'$ channel is quite different
from that of the $\sigma\to\pi'$ channel.
The curves in the $O_{xy}$, $O_{yz}$ and $O_{zx}$ phases can be
transformed into those for the $O_{yz}$, $O_{zx}$, and $O_{xy}$ phases,
respectively, by shifting $\psi$ with $2\pi/3$.
This threefold symmetry around ${\bf G}=(\frac{1}{2}\frac{1}{2}\frac{1}{2})$ 
perfectly matches the relation between the order parameters of three domains.
Thus the azimuthal-angle dependence is closely related to the geometry of 
scattering as well as the symmetry of the AFQ order, 
not related to the details of the model.
Therefore, the examination of the azimuthal angle dependence may be useful
to determine the symmetry of the AFQ phase.

\begin{figure}[t]
\vspace*{-0.6cm}
\centerline{\epsfxsize=7.50cm\epsfbox{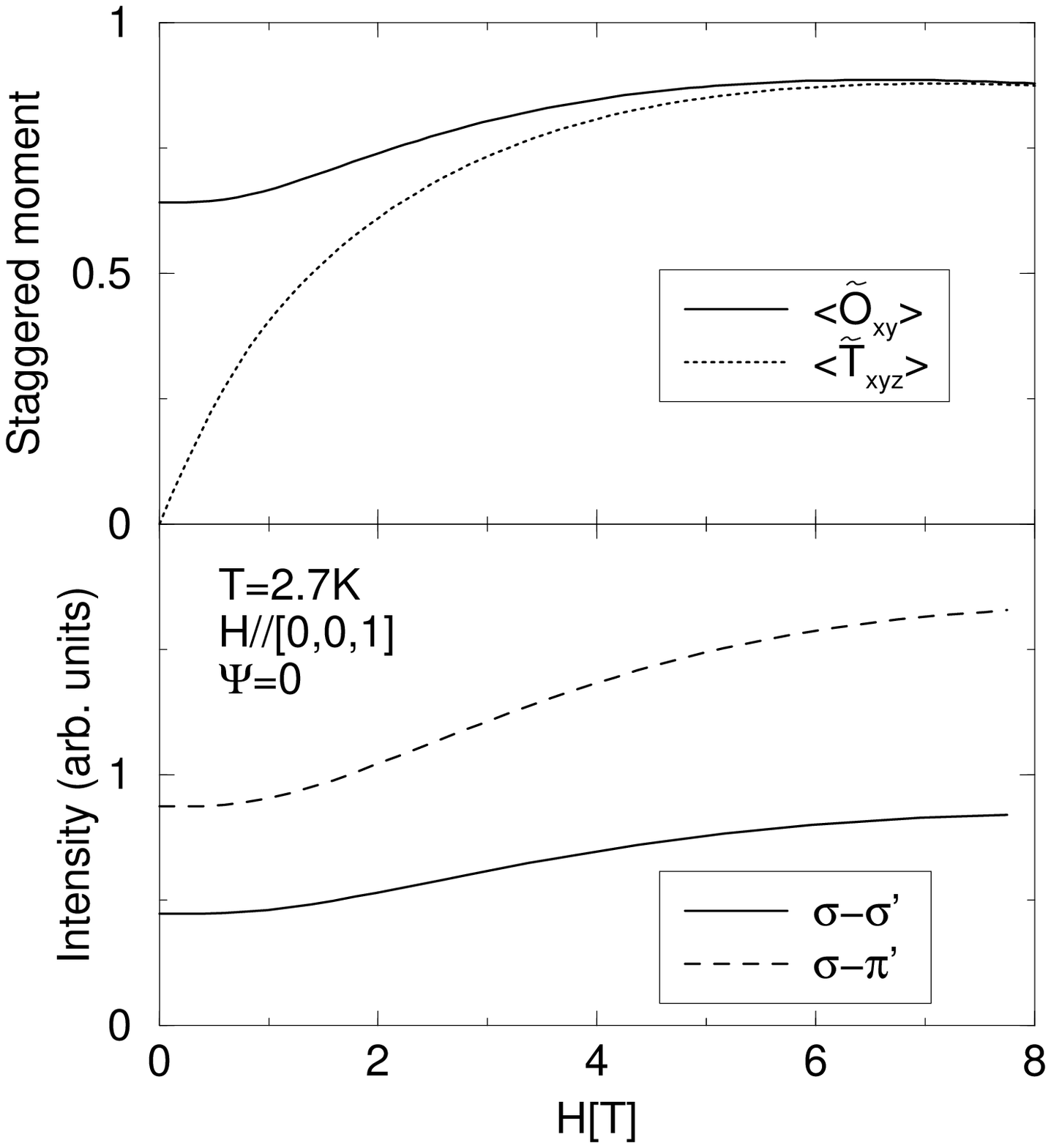}} 
\vskip 20pt
\caption{
RXS intensity of the peak at $\hbar\omega=5722$ eV
for ${\bf G}=(\frac{1}{2}\frac{1}{2}\frac{1}{2})$,
as a function of magnetic field.
${\bf H}\parallel [0,0,1]$, $T=2.7$ K and $\psi=0$.
The solid and broken lines represent the intensities for
the $\sigma\to\sigma'$ channel and the $\sigma\to\pi'$ channel,
respectively.
The upper panel shows the staggered quadrupole moment (solid line)
and the octupole moment (dotted line).
\label{fig.spext}}
\vspace*{-0.3cm}
\centerline{\epsfxsize=7.50cm\epsfbox{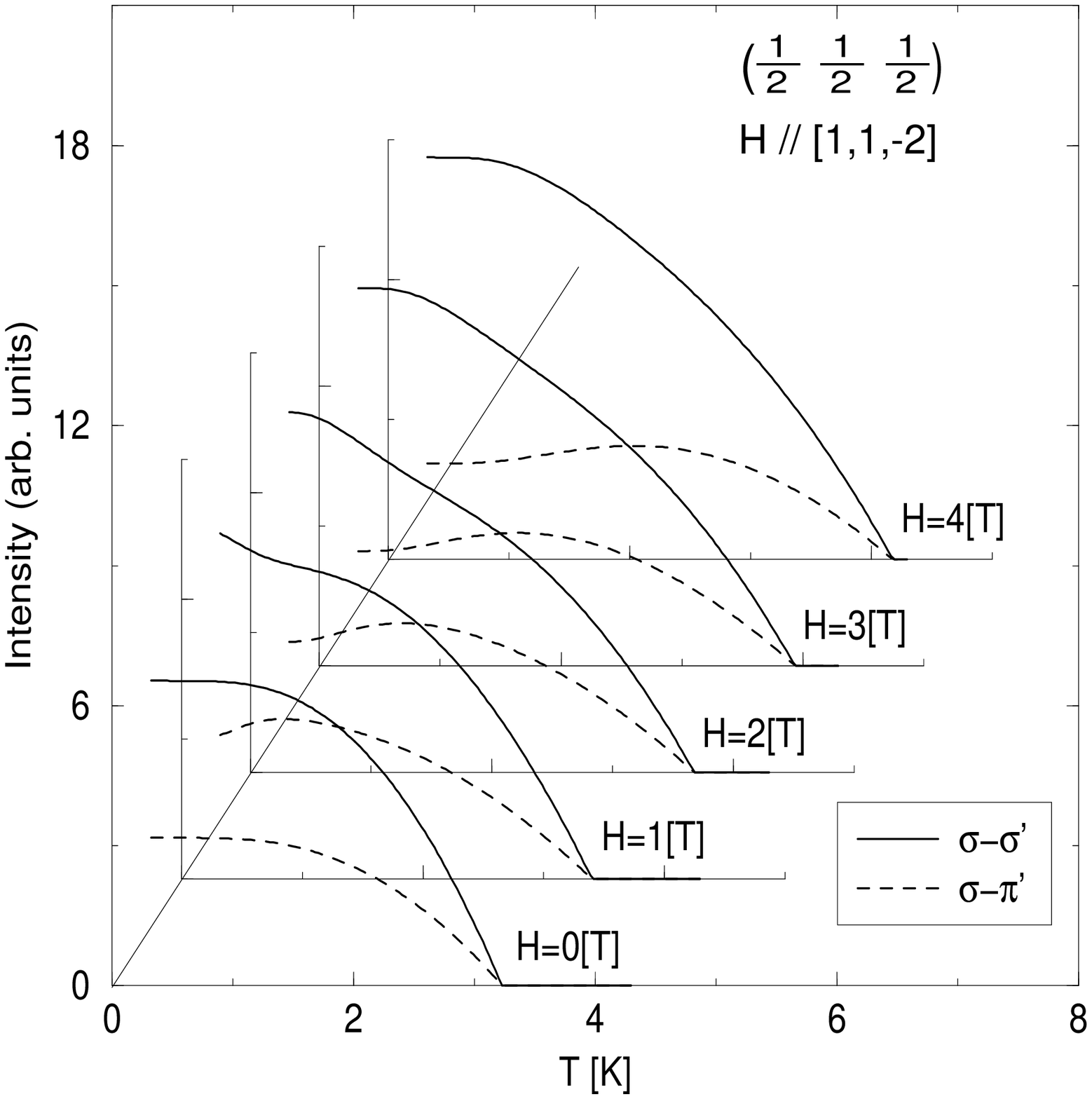}} 
\vskip 30pt
\caption{
RXS intensity of the peak at $\hbar\omega=5722$ eV as a function of 
temperature, for various values of $H$ (${\bf H}\parallel [1,1,-2]$).
${\bf G}=(\frac{1}{2}\frac{1}{2}\frac{1}{2})$.
The curve for $H=0$ is not the domain average but the limit of $H\to 0$.
\label{fig.sph112}}
\end{figure}

The initial state is rather sensitive to the magnetic field.
For example, as shown in the upper panel of Fig.~\ref{fig.spext},
a sizable staggered octupole moment $\langle\tilde T_{xyz}\rangle$ 
($\equiv 2\langle\tau^y\rangle$) is induced
by applying the magnetic field along the $[0,0,1]$ direction.
The lower panel of Fig.~\ref{fig.spext} shows the intensity of the peak 
at $\hbar\omega=5722$ eV as a function of magnetic field. 
The intensity increases only gradually with increasing $H$,
which behavior is close to the variation of the staggered quadrupole moment.
This indicates that the induced staggered octupole moment has little influence
on the RXS spectra. Finally in this subsection, we demonstrate 
in Fig.~\ref{fig.sph112} that
the calculated temperature dependence reproduces well
the experiment for ${\bf H}\parallel[1,1,-2]$.\cite{Nakao1} 

\subsection{Magnetic Ground State}

\begin{figure}[t]
\centerline{\epsfxsize=7.50cm\epsfbox{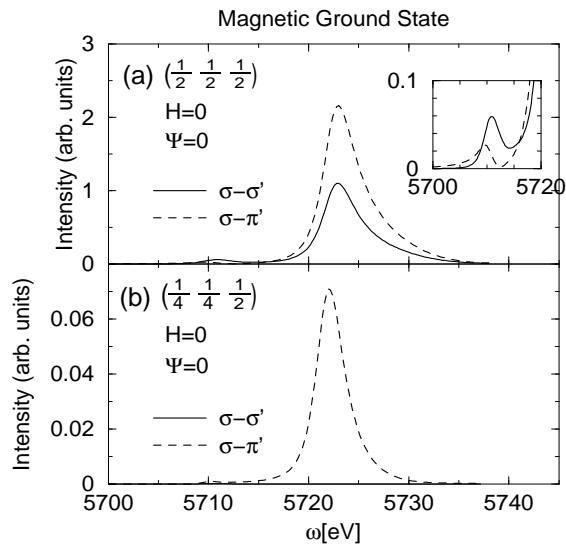}} 
\caption{
RXS spectra in the magnetic ground state as a function of photon energy:
(a) ${\bf G}=(\frac{1}{2}\frac{1}{2}\frac{1}{2})$;
(b) ${\bf G}=(\frac{1}{4}\frac{1}{4}\frac{1}{2})$.
$H=0$ and $\psi=0$. 
The pre-edge peak is magnified in the inset.
The solid and broken lines represent the intensities
for the $\sigma\to\sigma'$ channel and
the $\sigma\to\pi'$ channel, respectively.
\label{fig.spmag1}}
\end{figure}

Using the wave functions given in the preceding section for the magnetic
ground state, we calculate the RXS intensities for the AFQ 
superlattice spot ${\bf G}=(\frac{1}{2}\frac{1}{2}\frac{1}{2})$ and for
the magnetic superlattice spot ${\bf G}=(\frac{1}{4}\frac{1}{4}\frac{1}{2})$.
Figure \ref{fig.spmag1} shows the calculated spectra as a function of 
photon energy for $H=0$.

For ${\bf G}=(\frac{1}{2}\frac{1}{2}\frac{1}{2})$,
the spectral shape is close to the one in the AFQ phase.
The intensity of the main peak is a smooth extension from the $O_{xy}$ phase,
as shown in Fig.~\ref{fig.sptemp}.
This indicates that the primary origin is the anisotropic charge distribution 
associated with the AFQ order in the $4f$ states, and that
the magnetic order, which lifts the degeneracy of Kramers' doublet,
has little influence on this spot.
On the other hand, a pre-edge peak around $\hbar\omega=5710$ eV
is enhanced to become visible in the $E_2$ process.

For ${\bf G}=(\frac{1}{4}\frac{1}{4}\frac{1}{2})$, we have
the intensity two order of magnitude smaller than 
for ${\bf G}=(\frac{1}{2}\frac{1}{2}\frac{1}{2})$.
The origin is purely magnetic, that is, the main peak arises from 
the modulation of the $5d$ states in accordance with the magnetic order 
through the intra-atomic Coulomb interaction with the $4f$ states.
The contributions from modulation in accordance with the AFQ order 
are canceled out among different magnetic sublattices. 
No intensity comes out in the $\sigma\to\sigma'$ channel.
Note that the contribution of the non-resonant term 
(eq.~(\ref{eq.nonresonance})) is found too small to be seen.

\section{Concluding Remarks}

We have studied the RXS spectra near the Ce $L_{\rm III}$
absorption edge in the AFQ phase and also in the magnetic ground state,
on the basis of a microscopic model that the $2p$ and $4f$ states of Ce sites 
are atomic while the $5d$ states form an energy band 
with a reasonable DOS. 
Since the lattice distortion associated with the AFQ order has not been 
observed, we have paid no attention on it in the present calculation.
The $\Gamma_8$ states of the Ce $4f$ states have a lower energy than 
the $\Gamma_7$ states under the crystal field.

In the AFQ phase, we have employed an effective inter-site interaction 
between the $\Gamma_8$ states, which is developed by Shiina {\em et. al.},
and applied the MFA for describing the initial state.
In the description of the intermediate state, on the other hand,
those energies can be neglected in comparison with much larger energies
such as the intra-atomic Coulomb interaction and the energy of the $5d$ band.
Introducing the Green's function, and taking full account of the intra-atomic
Coulomb interaction, we have solved the scattering problem of the excited
$5d$ electron by the core-hole potential.
We have found that the initial state has a sensitive effect on the RXS spectra
through the dipole matrix elements. 

We have obtained relatively large RXS intensities on the AFQ superlattice spot
without assuming the lattice distortion,
thereby demonstrating the mechanism of the Coulomb interaction.
This situation is different from that of transition-metal compounds.
The temperature and magnetic field dependences of the RXS spectra 
reproduce well the experiment.\cite{Nakao1}
We have found that the azimuthal-angle dependence is 
closely related to the symmetry of the AFQ order.
Therefore this quantity may be useful to determine the order parameter.
We hope this study prompts experimentalists to measure this quantity.

In the magnetic ground state, we have constructed the wave function 
to be consistent with the ordering pattern of the magnetic moment 
determined by the neutron scattering experiment. 
Using this as the initial state, we have calculated
the RXS spectra on an AFQ superlattice spot and 
on a magnetic superlattice spot.
The intensity on the AFQ spot is a smooth extension from the AFQ phase.
In addition, we have found a small pre-edge peak in the $E_2$ process.
Since the main peak intensity will be reduced by the 
{\em absorption correction}, the pre-edge peak might be observed
in future experiment.
On the magnetic superlattice spot, we have a finite but much smaller 
intensity than that on the AFQ spot.
It may be hard to confirm experimentally the spectra on this spot.

We have assumed the shape of the $5d$-DOS rather arbitrarily.
This is partly justified by the fact that the characteristics
of the RXS spectra discussed in this paper do not sensitively depend
on the details of the $5d$ DOS.
Nevertheless, a band structure calculation for the $5d$ states 
may be necessary for more quantitative study.

\section*{Acknowledgements}
We would like to thank H. Nakao, Y. Murakami, and H. Shiba 
for valuable discussions.
This work was partially supported by 
a Grant-in-Aid for Scientific Research 
from the Ministry of Education, Science, Sports and Culture, Japan.


\begin{thebibliography}{99}
\def\vol(#1,#2,#3){{\bf #1} (#2) #3}
\bibitem{Murakami98a} Y. Murakami, H. Kawata, M. Tanaka, 
                      T. Arima, Y. Moritomo and Y. Tokura:
                      Phys. Rev. Lett. \vol(80,1998,1932). 

\bibitem{Murakami98b} Y. Murakami, J. P. Hill, D. Gibbs, M. Blume, I. Koyama,
                      M. Tanaka, H. Kawata, T. Arima, Y. Tokura,
                      K. Hirota and Y. Endoh:
                      Phys. Rev. Lett. \vol(81,1998,582). 

\bibitem{Murakami99b} K. Nakamura, T. Arima, A. Nakazawa, 
                      Y. Wakabayashi and Y.  Murakami:
                      Phys. Rev. B  \vol(60,1999,2425).   

\bibitem{Murakami99c} M. von Zimmermann, J.P. Hill, D. Gibbs, M. Blume, 
                      D. Casa, B. Keimer, Y. Murakami, Y. Tomioka 
                      and Y. Tokura:
                      Phys. Rev. Lett. \vol(83,1999,4872). 

\bibitem{Murakami00b} M. Noguchi, A. Nakazawa, T. Arima, Y. Wakabayashi, 
                      H. Nakao and Y. Murakami:
                      Phys. Rev. B \vol(62,2000,R9271).   

\bibitem{Ishihara1} S. Ishihara and S. Maekawa:
                    Phys. Rev. Lett. \vol(80,1998,3799).

\bibitem{Ishihara2} S. Ishihara and S. Maekawa:
                    Phys. Rev. \vol(B 58,1998,13449).

\bibitem{Elfimov} I. S. Elfimov, V. I. Anisimov and G. Sawatzky:
                  Phys. Rev. Lett. \vol(82,1999,4264).

\bibitem{Benfatto} M. Benfatto, Y. Joly and C. R. Natoli:
                   Phys. Rev. Lett. \vol(83,1999,636).

\bibitem{Takahashi1} M. Takahashi, J. Igarashi and P. Fulde:
                    J. Phys. Soc. Jpn. \vol(68,1999,2530).

\bibitem{Takahashi2} M. Takahashi, J. Igarashi and P. Fulde:
                    J. Phys. Soc. Jpn. \vol(69,2000,1614).

\bibitem{Takahashi3} M. Takahashi and J. Igarashi:
                     Phys. Rev. \vol(B 64,2001,075110).

\bibitem{Takahashi4} M. Takahashi and J. Igarashi:
                     unpublished.

\bibitem{Nakao1} H. Nakao, K. Magishi, Y. Wakabayashi, Y. Murakami,
                  K. Koyama, K. Hirota, Y. Endoh and S. Kunii:
                  J. Phys. Soc. Jpn. \vol(70,2001,1857).

\bibitem{Yakhou} F. Yakhou, V. Plakhty, H. Suzuki, S. Gavrilov,
                 P. Burlet, L. Paolasini, C. Vettier and S. Kunii:
                 Phys. Lett. \vol(A 285,2001,191).

\bibitem{Nagao} T. Nagao and J. Igarashi:
                J. Phys. Soc. Jpn. \vol(70,2001,2892).

\bibitem{Ohkawa1} F. J. Ohkawa:
               J. Phys. Soc. Jpn. \vol(52,1983,3897).

\bibitem{Ohkawa2} F. J. Ohkawa:
               J. Phys. Soc. Jpn.  \vol(54,1985,3909).

\bibitem{Shiina} R. Shiina, H. Shiba and P. Thalmeier:
                J. Phys. Soc. Jpn. \vol(66,1997,1741).

\bibitem{Sakai} O. Sakai, R. Shiina  H. Shiba and P. Thalmeier:
                J. Phys. Soc. Jpn. \vol(66,1997,3005).

\bibitem{Thalmeier} P. Thalmeier, R. Shiina, H. Shiba and O. Sakai:
                    J. Phys.Soc. Jpn. \vol(67,1998,2363).

\bibitem{Shiba} H. Shiba, O. Sakai and R. Shiina:
                J. Phys. Soc. Jpn.  \vol(68,1999,1988).

\bibitem{Effantin} J. M. Effantin, J. Rossat-Mignod, P. Burlet, H. Bartholin,
                   S. Kunii and T. Kasuya: 
                   J. Magn. Magn. Mater.  \vol(47\& 48,1985,145).

\bibitem{Takigawa} M. Takigawa, H. Yasuoka, T. Tanaka and Y. Ishizawa: 
                   J. Phys. Soc. Jpn. \vol(52,1983,3967).

\bibitem{Com1} In DyB$_2$C$_2$, a pre-edge peak is comparable to the main
               peak. See, Y. Tanaka, T. Inami, T. Nakamura, H. Yamauchi,
               H. Onodera, K. Ohyama and Y. Yamaguchi:
               J. Phys. Condens. Matter \vol(11,1999,L505);
               K. Hirota, N. Oumi, T. Matsumura, H. Nakao,
               Y. Wakabayashi, Y. Murakami and Y. Endoh:
               Phys. Rev. Lett. \vol(84,2000,2706).

\bibitem{deBergevin} F. de Bergevin and M. Brunel:
                     Acta Crystallogr., Sect. A: Cryst. Phys., Diffr.,
                     Theor. Gen. Crystallogr. \vol(37,1981,324). 

\bibitem{Blume1} M. Blume: J. Appl. Phys. \vol(57,1985,3615).

\bibitem{Blume2} M. Blume and D. Gibbs: Phys. Rev. B \vol(37,1988,1779).

\bibitem{Trammell} G. T. Trammell: Phys. Rev. \vol(92,1953,1387).

\bibitem{Cowan} R. Cowan: {\em The Theory of Atomic Structure and Spectra}
                (University of California Press, Berkeley, 1981).

\bibitem{Igarashi1} J. Igarashi and M. Takahashi:
                    J. Phys. Soc. Jpn. \vol(69,2000,4087).

\bibitem{Saitoh} M. Saitoh, H. Takagiwa, H. Ichikawa, T. Yokoo,
                 J. Akimitsu, M. Nishi, K. Kakurai, M. Takata,
                 N. Okada, M. Sakata and S. Kunii:
                 to be published in J. Phys. Soc. Jpn.

\bibitem{Hanzawa} K. Hanzawa: 
                  J. Phys. Soc. Jpn. \vol(70,2001,1900).
                  
\bibitem{Tsuji} S. Tsuji, M. Sera and K. Kojima:
                J. Phys. Soc. Jpn. \vol(70,2001,2864).

\bibitem{Com2} 
Explicitly, the $5d$-DOS, $\rho^{5d}(x)$, for each symmetry is given by
\[ \rho^{5d}(x) = \left\{ \begin{array}{ll}
                0.008x+0.04,    &  -5<x<0, \\
                0.01x+0.04,    &  0<x<8, \\
                -0.0277x+0.342,&  8<x<12.33, \\
  \end{array} \right.
\]
where $x$ is measured in units of eV with $x=0$ corresponding 
to the Fermi level.

\bibitem{Com3} 
It is known that the anisotropic terms of the Coulomb interaction
are slightly reduced in solids; we use the atomic values in Tables I
by reducing them with multiplying a factor 0.8. 
On the other hand, the values of $F^0(2p,4f)$, $F^0(2p,5d)$,
$F^0(4f,4f)$, and $F^0(4f,5d)$ are considerably screened in solids,
so that we use rather smaller values for them. 

\bibitem{Nakao2} H. Nakao, K. Magishi, Y. Wakabayashi, Y. Murakami,
                 K. Koyama, K. Hirota, Y. Endoh and S. Kunii:
                 to be published in J. Phys. Soc. Jpn.

\end{thebibliography}
\end{document}